\documentclass{article}

\PassOptionsToPackage{numbers, sort&compress}{natbib}
\usepackage[preprint]{neurips_2026}

\usepackage[utf8]{inputenc}
\usepackage[T1]{fontenc}
\usepackage{hyperref}
\usepackage{url}
\usepackage{booktabs}
\usepackage{amsfonts}
\usepackage{fancyvrb}
\usepackage{amsmath}
\usepackage{graphicx}
\usepackage{xcolor}
\usepackage{pifont} 
\usepackage{booktabs}
\usepackage{tikz}
\usetikzlibrary{arrows.meta, positioning, calc, backgrounds, fit, shapes.geometric, shapes.misc}

\title{Sovereign Agentic Loops: Decoupling AI Reasoning from Execution in Real-World Systems}

\author{
	Jun He \\
	OpenKedge.io \\
	\texttt{junhe@openkedge.io} \\
	\and 
	Deying Yu \\
	OpenKedge.io \\
	\texttt{deying@openkedge.io}
}

\begin{document}
	
	\maketitle
	
\begin{abstract}
Large language model (LLM) agents increasingly issue API calls that mutate real systems, yet many current architectures pass stochastic model outputs directly to execution layers. We argue that this coupling creates a safety risk because model correctness, context awareness, and alignment cannot be assumed at execution time. We introduce \textbf{Sovereign Agentic Loops (SAL)}, a control-plane architecture in which models emit structured intents with justifications, and the control plane validates those intents against true system state and policy before execution. SAL combines an obfuscation membrane, which limits model access to identity-sensitive state, with a cryptographically linked Evidence Chain for auditability and replay. We formalize SAL and show that, under the stated assumptions, it provides policy-bounded execution, identity isolation, and deterministic replay. In an OpenKedge prototype for cloud infrastructure, SAL blocks 93\% of unsafe intents at the policy layer, rejects the remaining 7\% via consistency checks, prevents unsafe executions in our benchmark, and adds 12.4 ms median latency.
\end{abstract}

\section{Introduction}

Large language models (LLMs) have accelerated interest in autonomous \emph{agentic systems} that reason, plan, and interact with external environments. By integrating LLMs with APIs that perform real-world actions---ranging from software deployment to infrastructure operations---these systems are moving from prototypes toward production use.

Despite this progress, many systems still rely on an architectural assumption that deserves closer scrutiny: model outputs are treated as executable commands. In these frameworks, an LLM observes a system state, generates an action payload, and that payload is dispatched with little or no mediation. This tightly coupled pipeline assumes that the reasoning process is sufficiently correct, context-aware, and aligned with system invariants at the moment of execution~\cite{lin2026safety}.

In practice, that assumption is unreliable. Unlike deterministic software components, frontier generative models are stochastic and often operate under partial observability or obfuscated context. Their failures are not limited to incorrect text generation; they can also produce unsafe real-world mutations. For example, an agent might issue a syntactically valid command such as \texttt{TerminateInstances} against a critical database node because it inferred the wrong operational context. Post-hoc safeguards such as logging or coarse permission boundaries may record or limit such behavior, but they do not by themselves verify whether the proposed action is appropriate under the true system state.

In classical control theory, unstable systems are not made safe by assuming ideal controller behavior; they are governed through feedback structures that constrain their effects~\cite{bode1945, astrom2010feedback}. Motivated by this perspective, we argue that stochastic reasoning should be bounded at the execution boundary rather than trusted to act directly on external systems.

We identify \emph{direct execution authority} as the core architectural problem. To address it, we formalize the \textbf{Decoupling Principle}: reasoning models should yield \emph{verifiable intent} rather than \emph{direct execution authority}. This reframes model outputs as proposals that require validation before they can affect system state.

We operationalize this principle through the \textbf{Sovereign Agentic Loop (SAL)}, a minimal-trust control-plane architecture that separates semantic reasoning from execution through a deterministic mediation pipeline. Under SAL, model generations are intercepted and evaluated as intent proposals against pre-execution policy constraints. 

To connect the formal framework to a deployable system, we implement SAL within the \textbf{OpenKedge}~\cite{openkedge2026} control plane. This implementation uses an \emph{obfuscation membrane} to limit model access to identity-sensitive information and a cryptographically linked \emph{Evidence Chain} to record states, intents, and executed actions.

Our core contributions are as follows:
\begin{itemize}
    \item We identify the coupling of probabilistic reasoning with direct execution authority as an architectural vulnerability in contemporary agentic systems.
    \item We formalize the \emph{Decoupling Principle} and introduce the \textbf{Sovereign Agentic Loop (SAL)}, a control-plane framework that evaluates model outputs as intents before execution.
    \item We define an \emph{obfuscation membrane} that isolates reasoning processes from identity-sensitive state while preserving the structural information needed for decision-making.
    \item We present an \emph{Evidence Chain} ledger that records intents, evaluations, and executed actions to support auditability and deterministic replay.
    \item We evaluate the architecture in the \textbf{OpenKedge} prototype, showing that unsafe actions can be blocked before execution with 12.4 ms median overhead in our benchmark.
\end{itemize}

Together, these contributions support a design approach in which agent safety is enforced by the execution architecture rather than delegated solely to model behavior.

\section{Related Work}

The problem of safely integrating autonomous agents with real-world systems intersects several research areas, including agentic tool use, runtime verification, access control, and control theory. We position Sovereign Agentic Loops (SAL) as a unifying control-plane abstraction that addresses limitations across these domains.

\subsection{Agentic Tool Use and LLM-Based Systems}

Recent advances in large language models (LLMs), including systems developed by OpenAI~\cite{openai2023gpt4} and Anthropic~\cite{anthropic2023claude}, have enabled agents to perform tool use and multi-step reasoning. Frameworks such as ReAct~\cite{yao2022react}, Toolformer~\cite{schick2023toolformer}, and function-calling APIs allow models to generate structured actions that interact with external systems. More recently, there has been a push towards formalized execution verification for LLM agents in production environments~\cite{wang2025agentic}, though many implementations remain constrained to read-only observability.

A recent systematic audit—the 2025 AI Agent Index~\cite{staufer2026agentindex}—documents the technical and safety features of 30 deployed agentic systems, revealing a \emph{transparency asymmetry}: while developers readily disclose capabilities, only four of 30 agents provide agent-specific safety evaluations, and 25 disclose no internal safety results whatsoever. This gap motivates structural execution governance.

These approaches focus primarily on improving the \emph{reasoning capabilities} of the model or the \emph{interface} between models and tools. However, they typically assume that once an action is generated, it can be executed directly, subject only to syntactic validation or lightweight guardrails~\cite{lin2026safety}.

In contrast, SAL introduces a structural decoupling between reasoning and execution. Rather than treating model outputs as executable commands, SAL treats them as \emph{intents} that must be validated within a sovereign control plane. This shifts the problem from tool orchestration to execution governance.

\subsection{Guardrails and Runtime Validation}

A growing body of work focuses on guardrails for LLMs, including prompt constraints, output filtering, rule-based validation layers~\cite{wei2023chainofthought, openai2023functioncalling}, and late-stage runtime monitors~\cite{lin2026safety}. These techniques aim to prevent unsafe outputs by constraining model behavior at inference time.

Recent work has advanced beyond static rule-based guardrails. AgentSpec~\cite{wang2025agentspec} introduces a domain-specific language for defining customizable runtime safety rules—specifying triggers, conditions, and enforcement mechanisms that are evaluated during agent execution. Pro2Guard~\cite{wang2025pro2guard} extends this line of work from reactive to \emph{proactive} enforcement, modeling agent behaviors as Discrete-Time Markov Chains (DTMCs) and using probabilistic reachability analysis to estimate the probability of reaching an unsafe state, enabling intervention before a violation materializes. Concurrently, the Open Agent Passport (OAP)~\cite{oap2026} provides a deterministic pre-action authorization specification that synchronously intercepts tool calls, evaluates them against declarative policies, and produces cryptographically signed audit records—achieving a 0\% adversarial success rate under restrictive policies with a median enforcement latency of 53~ms.

While effective for constraining specific agent behaviors, such approaches remain fundamentally \emph{model-centric} or \emph{action-centric}. They do not provide guarantees over the \emph{effects} of actions once executed in external systems, nor do they establish a complete causal chain linking reasoning to execution outcomes.

SAL differs in that safety is enforced \emph{prior to execution} through a deterministic evaluation function over real system state. The guarantees of SAL do not depend on the correctness or alignment of the underlying model, but on the enforcement properties of the control plane. While OAP shares SAL's pre-execution philosophy, SAL additionally formalizes the \emph{intent abstraction}, information-theoretic isolation via the obfuscation membrane, and cryptographic evidence chains as first-class primitives of the control plane.

\subsection{Runtime Verification and Formal Methods}

Runtime verification and formal methods provide mechanisms for checking system properties during execution~\cite{leucker2009brief, bartocci2018survey}. Techniques such as temporal logic monitoring, invariant enforcement, and model checking have been widely applied to distributed and safety-critical systems.

Recent work by Doshi et al.~\cite{doshi2026stpa} applies System-Theoretic Process Analysis (STPA) to LLM-based agents, systematically identifying hazards in agent workflows and deriving formal safety requirements for tool sequences and data flows. Their approach enhances the Model Context Protocol (MCP) with capability, confidentiality, and trust-level labels to enforce these requirements at runtime, moving from ad hoc reliability fixes toward formal safety contracts.

These approaches typically assume a known system model and operate over observable system traces. In agentic systems, however, the decision-making process is externalized to a stochastic model, and execution may be triggered without a formally verifiable decision boundary.

SAL complements runtime verification by introducing a \emph{pre-execution decision boundary}. Instead of verifying correctness after or during execution, SAL enforces policy compliance before any state mutation occurs. The Evidence Chain also provides a complete causal trace that enables deterministic replay, extending traditional trace-based verification. Unlike STPA-derived safety contracts that require design-time hazard enumeration, SAL enforces safety invariants dynamically through sovereign policy evaluation over live system state.

\subsection{Access Control and Policy Systems}

Access control mechanisms such as Role-Based Access Control (RBAC)~\cite{sandhu1996rbac}, Attribute-Based Access Control (ABAC)~\cite{hu2015abac}, and modern policy engines such as Open Policy Agent (OPA)~\cite{opa2019} and Cedar~\cite{cedar2023} provide structured ways to regulate system access.

\emph{Capability governance} extends static access control to agentic settings. Sidik and Rokach~\cite{sidik2026aethelgard} introduce Aethelgard, a framework that moves beyond static container-based sandboxing toward learned, adaptive capability governance. Aethelgard uses reinforcement learning to enforce minimum viable capability sets per task, dynamically scoping tool awareness to mitigate the \emph{capability overprovisioning problem}—where agents are routinely exposed to all available tools regardless of task requirements.

These systems operate on the assumption that the caller identity and request context are trustworthy representations of intent. Unlike traditional policy engines, SAL does not rely on trusted caller identity or static authorization rules. Instead, it derives execution authority dynamically from validated intent, real system state, and justification consistency within the control plane. In agentic systems, however, the caller (i.e., the agent) may generate actions based on incomplete or incorrect reasoning.

SAL extends access control by shifting from \emph{identity-based authorization} to \emph{intent-based authorization}. Execution authority is not statically assigned but dynamically derived from validated intent, context, and policy evaluation. While Aethelgard's learned capability scoping reduces the attack surface available to the agent, SAL adds a separate constraint: even within the permitted capability set, every proposed action must pass deterministic policy evaluation before any state mutation occurs.

\subsection{Multi-Agent Coordination and Failure Taxonomies}

The scaling of agentic systems beyond single-agent deployments introduces emergent coordination risks that compound the safety challenges addressed by SAL. Cemri et al.~\cite{cemri2025mast} introduce MAST, a taxonomy categorizing multi-agent LLM system failures into 14 distinct modes across specification issues, inter-agent misalignment, and task verification failures—analyzed over 1,600 execution traces. Their findings demonstrate that uncoordinated multi-agent systems can amplify errors by up to 17$\times$, while centralized architectures with validation bottlenecks contain amplification to approximately 4.4$\times$.

Hierarchical oversight architectures have been proposed to address these failure modes. Kim et al.~\cite{kim2025tao} present Tiered Agentic Oversight (TAO), a multi-agent framework that routes tasks to specialized agents based on complexity and risk, enabling higher tiers to oversee and correct lower-tier reasoning. Google Research~\cite{google2025scaling} provides a quantitative foundation for multi-agent scaling through an evaluation of 180 agent configurations, demonstrating that multi-agent coordination can boost performance by up to 80.9\% on parallelizable tasks while degrading performance by 39--70\% on sequential reasoning tasks.

SAL is orthogonal to and composable with multi-agent coordination architectures. While MAST taxonomizes failures and TAO introduces hierarchical oversight, neither provides a \emph{pre-execution invariant enforcement} layer that structurally prevents unsafe mutations regardless of the coordination topology. SAL can serve as the execution-boundary control plane for any agent within a multi-agent system, ensuring that the error amplification documented by Cemri et al. cannot propagate through actual infrastructure mutations.

\subsection{Control Theory and Feedback Systems}

Control theory models systems as feedback loops that regulate behavior based on observed state~\cite{astrom2010feedback}. Classical control systems assume deterministic dynamics and well-defined control inputs.

Agentic systems violate these assumptions due to stochastic reasoning and partial observability. SAL can be interpreted as a control-theoretic extension in which the controller operates over \emph{intent proposals} rather than direct control signals.

By introducing a closed-loop structure with evaluation, execution, and recording stages, SAL re-establishes controllability in systems where the decision-making process is otherwise opaque.

\subsection{Data Privacy and Information-Theoretic Isolation}

Techniques such as differential privacy~\cite{dwork2006calibrating}, secure enclaves, and data anonymization aim to limit information leakage when processing sensitive data. Recent explorations into sovereign AI infrastructure~\cite{kim2025sovereign} highlight the necessity of strict data boundaries when utilizing foreign foundation models.

SAL incorporates an information-theoretic perspective through the obfuscation membrane, which ensures that the reasoning model operates on a transformed state with no mutual information with identity-sensitive components. Unlike traditional privacy techniques that focus on data release, SAL applies isolation to the \emph{decision-making interface} itself.

\subsection{Summary}

Existing approaches address individual aspects of the problem—reasoning, validation, access control, or privacy—but do not provide a unified framework for governing autonomous execution. Recent advances in proactive guardrails~\cite{wang2025pro2guard}, runtime enforcement DSLs~\cite{wang2025agentspec}, system-theoretic safety analysis~\cite{doshi2026stpa}, and pre-action authorization~\cite{oap2026} represent important progress, yet each addresses a single facet of the problem without establishing an end-to-end execution governance architecture.

Anchored natively in the OpenKedge foundation, this work extends the OpenKedge control plane to contribute an abstraction that:

\begin{itemize}
    \item Decouples reasoning from execution authority
    \item Enforces safety as a pre-execution constraint
    \item Provides deterministic replay through evidence chains
    \item Establishes information-theoretic isolation between reasoning and infrastructure state
\end{itemize}

SAL is designed to be complementary to advances in model capability, runtime enforcement, multi-agent coordination, and system verification.

\section{System Model and Theoretical Foundations}
\label{sec:model}

\subsection{System Topology and State Representation}

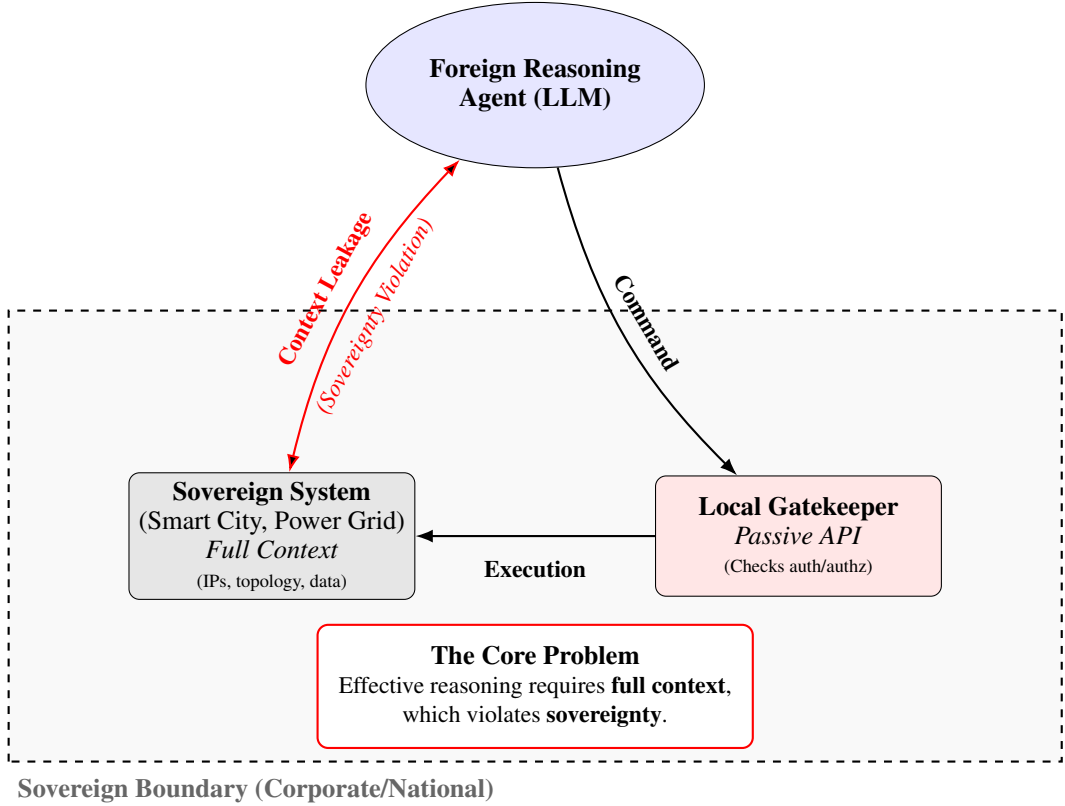
\begin{figure}[htpb]
\centering
\begin{tikzpicture}[
    >=Latex,
    box/.style={draw, rounded corners, minimum width=3.8cm, minimum height=1.6cm, align=center},
    cloud/.style={draw, ellipse, minimum width=4.5cm, minimum height=2.2cm, align=center},
    arrow/.style={->, thick},
    redarrow/.style={->, thick, draw=red}
]

\draw[dashed, thick, fill=gray!5] (-7,-3.5) rectangle (7,2.5);
\node[anchor=north west, text=gray!80!black] at (-7, -3.6) {\textbf{Sovereign Boundary (Corporate/National)}};

\node[box, fill=black!10] (system) at (-3.5, -0.5) {\textbf{Sovereign System} \\ (Smart City, Power Grid) \\ \textit{Full Context} \\ \scriptsize{(IPs, topology, data)}};
\node[box, fill=red!10] (gatekeeper) at (3.5, -0.5) {\textbf{Local Gatekeeper} \\ \textit{Passive API} \\ \scriptsize{(Checks auth/authz)}};

\node[cloud, fill=blue!10] (llm) at (0, 5.5) {\textbf{Foreign Reasoning} \\ \textbf{Agent (LLM)}};

\draw[<->, thick, draw=red] (system) edge[bend left=15] node[midway, sloped, above=1mm, font=\small, text=red] {\textbf{Context Leakage}} node[midway, sloped, below=1mm, font=\small, text=red] {\textit{(Sovereignty Violation)}} (llm);

\draw[arrow] (llm) edge[bend right=15] node[midway, sloped, above=1mm, font=\small] {\textbf{Command}} (gatekeeper);
\draw[arrow] (gatekeeper) -- (system) node[midway, below=2mm, font=\small] {\textbf{Execution}};

\node[draw=red, thick, fill=white, rounded corners, inner sep=8pt, align=center] (conflict) at (0, -2.5) {\textbf{The Core Problem} \\ \small Effective reasoning requires \textbf{full context}, \\ \small which violates \textbf{sovereignty}.};

\end{tikzpicture}
\caption{System topology illustrating the sovereignty conflict. A foreign reasoning agent requires full context from the sovereign system for effective decision-making, which leads to context leakage and a violation of data sovereignty.}
\label{fig:topology}
\end{figure}

As illustrated in Figure~\ref{fig:topology}, integrating foreign frontier models with sovereign infrastructure introduces a tension: the sovereign system encapsulates sensitive, localized context that must be protected, while the foreign reasoning agent requires this context to formulate effective decisions.

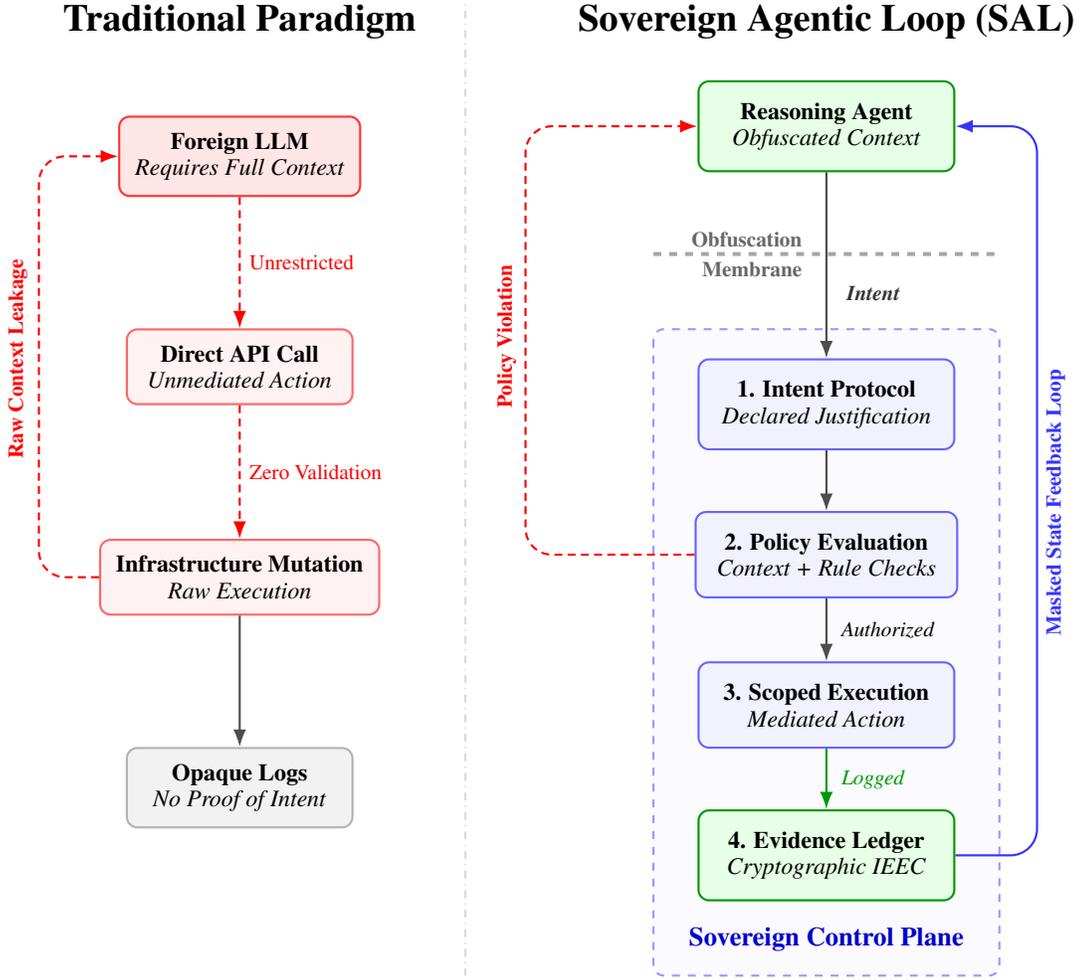
\begin{figure*}[t]
\centering
\begin{tikzpicture}[
    >=Latex,
    font=\small,
    tradnode/.style={draw=red!60, fill=red!5, rounded corners, inner sep=6pt, align=center, minimum width=3.0cm, thick},
    salnode/.style={draw=blue!60, fill=blue!5, rounded corners, inner sep=8pt, align=center, minimum width=3.4cm, thick},
    sysnode/.style={draw=gray!60, fill=gray!10, rounded corners, inner sep=6pt, align=center, minimum width=3.0cm, thick},
    arrow/.style={->, thick, draw=black!70},
    warnarrow/.style={->, thick, draw=red, densely dashed},
    passarrow/.style={->, thick, draw=green!60!black},
    looparrow/.style={->, thick, draw=blue!80, rounded corners=10pt}
]

\draw[dash dot, color=gray!30, thick] (0, 2.2) -- (0, -10.9);

\node at (-3.0, 1.8) {\Large \textbf{Traditional Paradigm}};

\node[sysnode, fill=red!10, draw=red!80] (trad_llm) at (-3.0, 0) {\textbf{Foreign LLM} \\ \textit{Requires Full Context}};
\node[tradnode] (trad_api) at (-3.0, -2.8) {\textbf{Direct API Call} \\ \textit{Unmediated Action}};
\node[tradnode] (trad_exec) at (-3.0, -5.6) {\textbf{Infrastructure Mutation} \\ \textit{Raw Execution}};
\node[sysnode] (trad_log) at (-3.0, -8.4) {\textbf{Opaque Logs} \\ \textit{No Proof of Intent}};

\draw[warnarrow] (trad_llm) -- node[right, font=\footnotesize, text=red] {Unrestricted} (trad_api);
\draw[warnarrow] (trad_api) -- node[right, font=\footnotesize, text=red] {Zero Validation} (trad_exec);
\draw[arrow] (trad_exec) -- (trad_log);

\draw[warnarrow, rounded corners=10pt] (trad_exec.west) -- ++(-0.8,0) |- (trad_llm.west);
\node[rotate=90, font=\footnotesize\bfseries, color=red, above=3pt] at (-5.7, -2.8) {Raw Context Leakage}; 

\node at (4.8, 1.8) {\Large \textbf{Sovereign Agentic Loop (SAL)}};

\node[salnode, fill=green!10, draw=green!60!black] (sal_llm) at (4.8, 0.4) {\textbf{Reasoning Agent} \\ \textit{Obfuscated Context}};

\draw[dashed, line width=1.5pt, color=gray!70] (2.5, -1.3) -- (7.1, -1.3);
\node[anchor=east, font=\footnotesize\bfseries, color=gray!80!black, align=right] at (4.6, -1.3) {Obfuscation\\[0.5ex]Membrane};

\draw[draw=blue!40, dashed, fill=blue!2, thick, rounded corners] (2.5, -2.3) rectangle (7.1, -10.9);
\node[anchor=south, font=\bfseries, color=blue!80!black] at (4.8, -10.7) {Sovereign Control Plane};

\node[salnode] (sal_intent) at (4.8, -3.3) {\textbf{1. Intent Protocol} \\ \textit{Declared Justification}};
\node[salnode] (sal_eval) at (4.8, -5.3) {\textbf{2. Policy Evaluation} \\ \textit{Context + Rule Checks}};
\node[salnode] (sal_exec) at (4.8, -7.3) {\textbf{3. Scoped Execution} \\ \textit{Mediated Action}};
\node[salnode, fill=green!10, draw=green!60!black] (sal_ieec) at (4.8, -9.3) {\textbf{4. Evidence Ledger} \\ \textit{Cryptographic IEEC}};

\draw[arrow] (sal_llm) -- node[right=4pt, pos=0.65, font=\footnotesize\bfseries, color=black!80] {\textit{Intent}} (sal_intent);
\draw[arrow] (sal_intent) -- (sal_eval);
\draw[arrow] (sal_eval) -- node[right=2pt, font=\footnotesize] {\textit{Authorized}} (sal_exec);
\draw[passarrow] (sal_exec) -- node[right=2pt, font=\footnotesize, text=green!60!black] {\textit{Logged}} (sal_ieec);

\draw[warnarrow, rounded corners=10pt] (sal_eval.west) -- (0.8, -5.3) -- (0.8, 0.4) -- (sal_llm.west);
\node[rotate=90, font=\footnotesize\bfseries, color=red, above=3pt] at (0.8, -2.5) {Policy Violation};

\draw[looparrow] (sal_ieec.east) -- (7.6, -9.3) -- (7.6, 0.4) -- (sal_llm.east);
\node[rotate=90, font=\footnotesize\bfseries, color=blue!80, below=3pt] at (7.6, -4.5) {Masked State Feedback Loop};

\end{tikzpicture}
\caption{Comparison between traditional agent execution and the Sovereign Agentic Loop (SAL). In the traditional model (left), actions are directly executed with no validation, resulting in opaque and non-replayable behavior. In SAL (right), all actions cross the Obfuscation Membrane into a sovereign control plane, enforcing policy validation, execution scoping, and cryptographic evidence recording.}
\label{fig:sal_comparison}
\end{figure*}

We formalize the operating environment of a sovereign agentic system, as depicted in Figure~\ref{fig:sal_comparison}, through a bisected state-space representation. This consists of the \textbf{true state space} $\mathcal{S}$, which contains identity-sensitive attributes, and the \textbf{obfuscated state space} $\hat{\mathcal{S}}$, a synthesized projection that preserves the structural topology required for higher-order reasoning while stripping identifying contexts.

We mathematically model any given state $s \in \mathcal{S}$ via a structural decomposition:
\begin{equation}
s = (s_{id}, s_{struct})
\end{equation}
where $s_{id}$ encompasses the identity-sensitive components (e.g., exact IP addresses, node IDs), and $s_{struct}$ denotes the structural properties.

\paragraph{Obfuscation Membrane.} To enforce information-theoretic isolation, we introduce the \textbf{obfuscation membrane}, formalized as a projection mapping $\Pi$:
\begin{equation}
\Pi: \mathcal{S} \rightarrow \hat{\mathcal{S}}
\end{equation}
Treating the system state as a random variable $S = (S_{id}, S_{struct})$ and defining $\hat{S} = \Pi(S)$, this projection must satisfy the strict isolation constraint:
\begin{equation}
I(S_{id}; \hat{S}) = 0
\end{equation}
where $I(\cdot; \cdot)$ denotes Shannon mutual information. This constraint implies that the identifying sub-state is not recoverable from the obfuscated representation under the model assumptions. In practice, this isolation holds provided that auxiliary side channels do not reintroduce identity information. 

Concurrently, the obfuscation must optimize for structural preservation to maintain reasoning efficacy:
\begin{equation}
\max_{\Pi} \; I(S_{struct}; \hat{S})
\end{equation}
Together, these constraints isolate identifying variables while preserving the spatial geometry needed for remote agentic reasoning.

To reverse this mapping within a secured boundary, we define a privileged reconstruction function $\Pi^{-1}$, accessible exclusively within the control plane:
\begin{equation}
\Pi^{-1}: \hat{\mathcal{S}} \times \mathcal{K} \rightarrow \mathcal{S}
\end{equation}
where $\mathcal{K}$ represents a sovereign mapping key or cryptographic lookup structure.

\subsection{The Sovereign Agentic Loop Formulation}

\paragraph{Agentic Reasoning Function.} Let $M$ denote a foreign reasoning model (e.g., a frontier LLM). The model strictly receives obfuscated inputs to generate formal intent and natural language justification:
\begin{equation}
M(\hat{s}, G) \rightarrow (i, J)
\end{equation}
where $\hat{s} \in \hat{\mathcal{S}}$ is the obfuscated state context, $G$ is the goal, $i \in \mathcal{I}$ is the structured intent, and $J$ acts as an explanatory reasoning trace. Consequently, the model's conditional generation carries zero mutual information regarding local infrastructure identity:
\begin{equation}
I\big(S_{id}; M(\hat{S}, G)\big) = 0
\end{equation}
In practice, this isolation holds under the assumption that external signals do not reintroduce identity information.

\paragraph{Unstable Reasoning Model.} We formally define the agent $M$ as a stochastic, \emph{unstable decision process}. For any given state-goal pair, the probability distribution over generated intents is formally non-stationary:
\begin{equation}
\exists \hat{s}, G, \; i_1 \neq i_2 \quad \text{such that} \quad 
P\big(M(\hat{s}, G) = (i_1, J_1)\big) > 0 \;\land\; P\big(M(\hat{s}, G) = (i_2, J_2)\big) > 0
\end{equation}
This non-determinism means that identical or near-identical scenarios can yield conflicting intents. This motivates treating model outputs as proposals that require downstream validation rather than as commands that can be executed directly.

\paragraph{Sovereign Evaluation Function.} To govern the stochastic generated intents, we define a deterministic control-plane evaluation function:
\begin{equation}
E: \mathcal{I} \times \mathcal{S} \times \mathcal{P} \rightarrow \{0,1\}
\end{equation}
where $\mathcal{P}$ represents the set of invariant sovereign policies. We logically decompose $E$ to validate both rule-based policy compliance and contextual reasoning alignment:
\begin{equation}
E(i, s, \mathcal{P}) = E_{\mathrm{policy}}(i, s, \mathcal{P}) \land E_{\mathrm{consistency}}(i, J, \hat{s})
\end{equation}
Here, $E_{\mathrm{policy}}: \mathcal{I} \times \mathcal{S} \times \mathcal{P} \rightarrow \{0,1\}$ validates compliance against system constraints using the full state $s$, while $E_{\mathrm{consistency}}: \mathcal{I} \times \mathcal{J} \times \hat{\mathcal{S}} \rightarrow \{0,1\}$ checks whether the proposed intent is consistent with its justification and the perceived context $\hat{s}$.

\paragraph{Execution Semantics.} Once an intent is authorized, the system executes it via a strictly mediated execution operator $\mathcal{X}$:
\begin{equation}
\mathcal{X}: \mathcal{I} \times \mathcal{S} \times \mathcal{P} \rightarrow \mathcal{A} \cup \{\bot\}
\end{equation}
where physical execution is explicitly conditioned on the evaluation ruling:
\begin{equation}
\mathcal{X}(i, s, \mathcal{P}) =
\begin{cases}
a_{\mathrm{exec}} & \text{if } E(i, s, \mathcal{P}) = 1 \\
\bot & \text{otherwise}
\end{cases}
\end{equation}
where the action $a_{\mathrm{exec}} = \Pi^{-1}(i)$ uses the privileged reconstruction mapping $\mathcal{K}$ to translate the abstract intent into a concrete mutation in the true infrastructure state space.

\paragraph{Evidence Chain.} To support auditability, we formalize the \emph{Evidence Chain} $\mathcal{C}$ as a cryptographically linked execution ledger. At step $t$, the chain records:
\begin{equation}
\mathcal{C}_t = \langle i_t, J_t, E_t, a_{\mathrm{exec}, t}, H(c_t), H(\mathcal{C}_{t-1}) \rangle
\end{equation}
where $a_{\mathrm{exec}, t} = \Pi^{-1}(i_t)$ (if authorized and executed), $c_t \subseteq s_t$ captures the precise context snapshot used during evaluation, and $H(\cdot)$ denotes a collision-resistant cryptographic hash function.

\subsection{Formal Safety Guarantees}

\paragraph{Assumptions.}
We adopt the following assumptions for the SAL model:

\begin{itemize}
    \item \textbf{A1 (Deterministic Evaluation):} The evaluation function $E$ produces a consistent decision for a given input $(i, s, \mathcal{P})$.
    \item \textbf{A2 (Deterministic Execution):} The execution operator $\mathcal{X}$ deterministically maps an approved intent to a concrete action.
    \item \textbf{A3 (Closed Execution):} All state mutations are mediated exclusively by $\mathcal{X}$.
    \item \textbf{A4 (Context Completeness):} The context $c_t$ contains all information required to evaluate the correctness and safety of an intent.
\end{itemize}

In practice, the closed execution assumption (A3) is enforced by restricting all mutation-capable credentials to the control plane, ensuring that no external component can invoke execution APIs directly.

\textbf{Lemma 1 (Non-Invertibility Without Sovereign Key).} \textit{Let $\Pi : \mathcal{S} \rightarrow \hat{\mathcal{S}}$ be the obfuscation membrane satisfying $I(S_{id}; \hat{S}) = 0$. Then, without access to the privileged mapping key $\mathcal{K}$ (or an equivalent reconstruction oracle), there exists no deterministic estimator $f$ such that:
\begin{equation}
f(\hat{S}) = S_{id}
\end{equation}
with an accuracy exceeding that of uninformed random guessing over the prior marginal distribution of $S_{id}$.}

\textit{Proof Sketch.}
Given the obfuscation constraint $I(S_{id}; \hat{S}) = 0$, the obfuscated state $\hat{S}$ shares no mutual information with the identifying attributes $S_{id}$. Consequently, an estimator $f(\hat{S})$ cannot improve on the marginal prior $P(S_{id})$ using only $\hat{S}$. Structural inversion therefore requires auxiliary side-channel information not contained in $\hat{S}$. In SAL, this information is confined to the privileged state-cache $\mathcal{K}$ within the control plane.

These guarantees hold strictly under the closed execution assumption (A3) and do not extend to systems with external execution pathways.\hfill $\square$

\textbf{Theorem 1 (Execution-Bound Sovereignty).} \textit{Assume a system satisfies the topological definitions established in Section~\ref{sec:model} and operates under assumptions A1–A4. Then, for every actualized execution $a_{\mathrm{exec}, t}$, the SAL architecture structurally guarantees:}

\begin{enumerate}
    \item \textbf{Policy-Bounded Execution:}
    \begin{equation}
    \forall a_{\mathrm{exec}, t}, \; \exists i_t \text{ such that } E(i_t, s_t, \mathcal{P}) = 1
    \end{equation}
    
    \item \textbf{Identity Isolation:}
    \begin{equation}
    I(A_{\mathrm{exec}, t}; S_{id} \mid \hat{S}_t) = 0
    \end{equation}
    
    \item \textbf{Deterministic Cryptographic Replay:}
    \begin{equation}
    \mathrm{Replay}(\mathcal{C}_t) = a_{\mathrm{exec}, t}
    \end{equation}
\end{enumerate}

\textbf{Proof Sketch.}
First, by the closed execution assumption (A3), the execution operator $\mathcal{X}$ mediates all possible state mutations. As formalized in the execution semantics, $\mathcal{X}$ yields an actionable mutation if and only if $E(i_t, s_t, \mathcal{P}) = 1$, which bounds execution to policy-approved intents.

Second, identity isolation follows from the obfuscation projection $I(S_{id}; \hat{S}_t) = 0$. Since the agent $M$ computes intent using $\hat{S}_t$ and $G$, the resulting $i_t$ is generated without direct access to identifying state. The mapping from $i_t$ to $a_{\mathrm{exec}, t}$ occurs within the control plane through $\Pi^{-1}$.

Third, deterministic replay follows from assumptions A1 and A2. Because the Evidence Chain ledger $\mathcal{C}_t$ records $i_t, c_t, E_t$, and the chained hash state $H(\mathcal{C}_{t-1})$, the information required to reconstruct the executed mutation remains available for replay and audit.

Consequently, the architecture guarantees policy-bounded execution, identity isolation, and deterministic replayability, completing the proof.\hfill $\square$

\paragraph{Summary.} Theorem 1 shows that, under the stated assumptions, safety properties can be enforced \emph{prior} to physical execution by the control plane rather than by relying on model behavior alone. This formulation establishes a closed-loop pipeline from obfuscated observation to intent generation, evaluation, and execution, forming the basis of execution-bound safety guarantees.

\section{System Design and Implementation}

We implement the Sovereign Agentic Loop (SAL) in a prototype system called \textbf{OpenKedge}~\cite{openkedge2026}, designed to integrate external reasoning agents with cloud infrastructure environments under control-plane mediation. OpenKedge intercepts, translates, evaluates, and governs agent-initiated mutations before they reach infrastructure APIs.

\subsection{Architectural Principles}

The OpenKedge implementation is organized around four design principles that operationalize the theoretical bounds established in Section~\ref{sec:model}:
\begin{enumerate}
    \item \textbf{Execution Sovereignty:} Infrastructure mutations pass through the control plane rather than directly from the reasoning model to the execution layer.
    \item \textbf{Minimal Trust in Reasoning:} The reasoning model is treated as a stochastic component whose outputs require validation before execution.
    \item \textbf{Deterministic Replayability:} Executed mutations should be reproducible from an immutable ledger.
    \item \textbf{Pluggable Enforcement:} Evaluation engines must decouple from execution layers, allowing integration with existing enterprise rule sets (e.g., OPA, AWS IAM) without disrupting structural safety.
\end{enumerate}

To realize these guarantees, the system is decomposed into six stateless modular components: an \textbf{Intent Interface} to normalize inputs, an \textbf{Obfuscation Membrane} to limit semantic leakage, a \textbf{Context Aggregator} to synthesize runtime metrics, a deterministic \textbf{Evaluation Engine}, an environment-specific \textbf{Execution Adapter}, and an append-only \textbf{Evidence Ledger}. Stateless operation localizes persistence to the evidence log and simplifies auditing.

\subsection{State Obfuscation and Intent Abstraction}

\paragraph{Obfuscation Membrane.} The implementation of the mapping projection $\Pi$ replaces sensitive production topology elements with structurally equivalent abstract graphs. OpenKedge strips identifying metadata---such as AWS account IDs, exact subnets, routing hashes, and internal DNS endpoints---and replaces them with ephemeral tokens (e.g., \texttt{Node-A7x}). The dependency graph topology is preserved so that the model can still reason over structural relationships. $\Pi$ and its inverse $\Pi^{-1}$ are maintained in a secure in-memory mapping cache within the control plane.

\paragraph{Normalized Intent Representation.} Outputs generated by the reasoning model are intercepted and parsed into a structured intent format $\mathcal{I}$. OpenKedge enforces the schema:
\begin{equation}
i = \langle \text{action\_type}, \text{target\_token}, \text{parameters}, \text{justification} \rangle
\end{equation}
For example, an intent to remove an anomalous compute instance is represented as a structured tuple rather than a raw shell command:
\begin{quote}
\texttt{action: TERMINATE\_NODE, target: Node-A7x, justification: "Memory exhaustion"}
\end{quote}
By requiring the agent to communicate through this intent protocol, OpenKedge separates semantic reasoning from executable commands and reduces the opportunity for prompt-injected script execution or direct API payload generation.

\subsection{Deterministic Sovereign Evaluation}

The OpenKedge evaluation engine deterministically validates an intent $i$ using the localized true state $s$ against predefined policies $\mathcal{P}$. The function $E(i, s, \mathcal{P})$ executes via two distinct pipelines:

\paragraph{Policy Alignment.} OpenKedge uses Open Policy Agent (OPA) to evaluate infrastructure mutations against explicit safety boundaries, such as production-destruction limits, blast-radius thresholds, and region-specific workload restrictions. Any intent that violates these rules evaluates to zero ($E_{\mathrm{policy}} = 0$) and is rejected.

\paragraph{Contextual Consistency.} Policy validation alone is insufficient if the intent's justification contradicts the observed system state. The consistency pipeline ($E_{\mathrm{consistency}}$) compares the generated reasoning trace $J$ with the real-time context $c_t$ pulled from the Aggregator. For example, if an agent proposes terminating a node because of "high CPU utilization" while telemetry shows nominal 15\% load, the consistency check fails.

\subsection{Mediated Execution and Evidence Chaining}

\paragraph{Transient Execution Domains.} Authorized intents that cross the evaluation threshold ($E = 1$) are translated and dispatched to environment-specific execution adapters (e.g., an AWS execution layer). To reduce the risk of persistent unauthorized access, the execution operator $\mathcal{X}$ provisions short-lived, scoped Identity and Access Management (IAM) tokens. These credentials expire upon action completion and tie execution privileges to the authorization window of the validated intent.

\paragraph{Cryptographic Evidence Ledger.} Upon completion, the precise decision lifecycle is symmetrically hashed into the immutable Evidence Chain ledger:
\begin{equation}
\mathcal{C}_t = \langle i_t, J_t, E_t, a_{\mathrm{exec}, t}, H(c_t), H(\mathcal{C}_{t-1}) \rangle
\end{equation}
This ledger supports deterministic replay, programmatic auditability, and forensic inspection of system mutations. 

\paragraph{Fail-Safe Resilience.} OpenKedge follows a default-deny fallback model. If action parsing fails, a context check cannot be completed, or an adapter faults, processing terminates in a denied state recorded in the chain rather than falling back to unmediated operations. This design keeps execution authority inside the control plane even when upstream reasoning or downstream infrastructure components fail.

\section{Evaluation}

We evaluate OpenKedge along two dimensions: (1) the latency overhead introduced by the Sovereign Agentic Loop (SAL) and (2) the empirical effectiveness of the architecture in preventing unsafe infrastructure mutations.

\subsection{Experimental Setup}

We deployed the OpenKedge prototype on a high-throughput testbed utilizing AWS EC2 \texttt{c6g.2xlarge} instances (AWS Graviton2, 8 vCPUs, 16GB RAM) running Amazon Linux 2026. The \emph{Obfuscation Membrane} utilized an isolated Redis cluster for in-memory token mapping ($\mathcal{K}$), while the \emph{Evidence Ledger} applied SHA-256 for cryptographic chaining. The deterministic policy engine was implemented using open-source Open Policy Agent (OPA) binaries.

To evaluate the control plane under varied conditions, we curated a dataset of $N_{\text{total}} = 10,000$ intent evaluation traces targeting simulated cloud infrastructure operations (e.g., EC2 termination, IAM role escalation, and VPC topology shifts). This workload was divided into:
\begin{itemize}
    \item \textbf{Benign Workload ($N_{\text{valid}} = 7,500$):} Routine, correctly structured operational requests representing compliant agent behavior.
    \item \textbf{Adversarial Workload ($N_{\text{unsafe}} = 2,500$):} Context-inconsistent or unsafe intents generated by red-teaming a simulated frontier model.
\end{itemize}

\subsection{Latency Overhead}

A key concern for inline mediation is throughput degradation. We profiled the end-to-end evaluation latency introduced by the control plane, measuring the time elapsed from intent ingestion to final Evidence Ledger serialization. Timing profiles were collected using high-resolution monotonic clocks across the $10,000$ simulated traces.

The measured overhead exhibited a median latency of:
\begin{equation}
\Delta t_{\text{median}} \approx 12.4 \text{ ms}
\end{equation}
with a tightly constrained 95\textsuperscript{th} percentile of $\Delta t_{95} \approx 21.7 \text{ ms}$. 

By instrumenting the specific subsystems, we decomposed the median 12.4 ms latency burden into:
\begin{enumerate}
    \item \textbf{De-obfuscation and Context Fetch ($\Pi^{-1}$ extraction):} $2.1 \text{ ms}$
    \item \textbf{OPA Policy Evaluation ($E_{\mathrm{policy}}$):} $6.8 \text{ ms}$
    \item \textbf{Consistency Evaluation ($E_{\mathrm{consistency}}$):} $2.3 \text{ ms}$
    \item \textbf{Cryptographic Hashing \& Ledger Serialization:} $1.2 \text{ ms}$
\end{enumerate}

This result suggests that intent mediation adds modest overhead relative to typical network jitter and is compatible with cloud orchestration workloads.

\subsection{Safety Effectiveness}

To quantify the protective capability of the architecture, we evaluated how the system handled the 2,500 adversarial intents ($N_{\text{unsafe}}$). 

Of the adversarial intents, OpenKedge successfully detected and blocked 2,325 (93.0\%) purely at the stateless rule-based layer ($E_{\mathrm{policy}} = 0$). These intents attempted direct violations of invariant boundaries, such as deleting active production databases or bypassing cross-region restrictions. 

The remaining 175 adversarial intents (7.0\%) satisfied the policy constraints but contained semantic contradictions (e.g., proposing an emergency cluster eviction because of "critical high temperature" while the aggregated telemetry graph indicated nominal temperature states). These attempts were rejected by the consistency evaluator ($E_{\mathrm{consistency}} = 0$). 

The combined blocking rate was therefore:
\begin{equation}
\text{Blocking Rate} = \frac{N_{\text{blocked\_policy}} + N_{\text{blocked\_consistency}}}{N_{\text{unsafe}}} = \frac{2325 + 175}{2500} = 1.00
\end{equation}
In this benchmark, no unsafe intent bypassed the evaluation stage to execute on the underlying infrastructure.

\subsection{Replay Verification}

Auditability is an important requirement for mediated execution. To validate the deterministic replay property, we used an isolated verification script to reconstruct the logic for all $7,500$ successfully executed benign actions by traversing the \emph{Evidence Chain}.

For every conditionally executed mutation within the testbed, the deterministic reconstruction yielded:
\begin{equation}
\text{Replay}(\mathcal{C}_t) = a_{\mathrm{exec}, t}
\end{equation}
yielding a 100\% verification match rate across the dataset, with no observed cryptographic desynchronization. This indicates that the contextual input ($c_t$) and justification ($J_t$) needed to reconstruct each policy decision were preserved in the chain.

\subsection{Summary}

The empirical evaluation is consistent with the theoretical framework. In our benchmark, the control plane blocked unsafe intents before execution while adding roughly 12.4 ms median latency. These results suggest that mediated execution can improve operational safety without imposing large runtime cost.

\section{Conclusion}

Agentic execution is becoming increasingly relevant in cloud and infrastructure systems. However, many current architectures still grant stochastic reasoning models direct access to execution interfaces, creating a pathway from incorrect or unsafe generations to real system mutations.

We introduced the \emph{Sovereign Agentic Loop} (SAL), a control-plane architecture that enforces \textbf{Execution Decoupling}. By separating semantic reasoning from the authority to mutate system state, SAL routes agent intents through a mediation pipeline before execution. We formalized this framework and showed that, under the stated assumptions, it provides policy-bounded execution, identity isolation via obfuscation, and auditable replay through the Evidence Chain.

In our empirical evaluation using the OpenKedge prototype, SAL blocked unsafe intents before execution in the benchmark workload. The mediation pipeline---including obfuscation, OPA policy evaluation, contextual consistency checks, and cryptographic chaining---added approximately $12.4\text{ ms}$ median latency. These results indicate that execution mediation can be practical for low-latency automation settings.

From a control-theoretic perspective, frontier AI agents introduce a stochastic decision process into operational domains that demand reliable behavior. Rather than relying solely on prompt design or alignment to eliminate that variability, SAL constrains its effects at the execution boundary. In this sense, SAL follows the engineering principle of managing instability through system structure~\cite{bode1945}.

As AI systems are deployed across organizational and regulatory boundaries, the case for mediated execution becomes stronger. SAL suggests one path toward that goal: treat model outputs as proposals, validate them against system state and policy, and execute only after that decision boundary has been crossed.

\bibliographystyle{unsrtnat}
\bibliography{references}

\appendix
\section{Notation}
For clarity, we summarize the notation used throughout the paper.
\begin{table}[h]
\centering
\small
\begin{tabular}{ll}
\toprule
\textbf{Symbol} & \textbf{Description} \\
\midrule

$\mathcal{S}$ & True state space of the infrastructure \\
$s \in \mathcal{S}$ & Concrete system state \\
$\hat{\mathcal{S}}$ & Obfuscated state space \\
$\hat{s} \in \hat{\mathcal{S}}$ & Obfuscated system state \\

$s_{id}$ & Identity-sensitive component of the state \\
$s_{\text{struct}}$ & Structural (topological) component of the state \\

$\Pi$ & Obfuscation mapping $\Pi: \mathcal{S} \rightarrow \hat{\mathcal{S}}$ \\
$\Pi^{-1}$ & Privileged reconstruction mapping within control plane \\

$\mathcal{I}$ & Intent space \\
$i \in \mathcal{I}$ & Structured intent generated by the agent \\

$M$ & Reasoning model (e.g., LLM) \\
$G$ & Goal specification provided to the agent \\
$J$ & Justification generated by the reasoning model \\

$E$ & Evaluation function for intent validation \\
$\mathcal{P}$ & Set of sovereign policy constraints \\

$\mathcal{X}$ & Execution operator mapping intent to action \\
$a$ & Abstract action \\
$a_{\mathrm{exec}}$ & Executed action in the real system \\

$\mathcal{C}_t$ & Evidence chain at time $t$ \\
$c_t$ & Context snapshot used for evaluation \\
$H(\cdot)$ & Cryptographic hash function \\

$I(\cdot;\cdot)$ & Mutual information between random variables \\

\bottomrule
\end{tabular}
\caption{Summary of notation used in the SAL model.}
\label{tab:notation}
\end{table}
	
\section{Figure-to-Notation Mapping}

To clarify the correspondence between the conceptual architecture illustrated in Figure~\ref{fig:sal_comparison} and the formal notation introduced in Section~\ref{sec:model}, we provide the following mapping. We explicitly align the conceptual components in Figure~\ref{fig:sal_comparison} with the formal model to ensure consistency between the architectural and mathematical representations.

\begin{table}[h]
\centering
\small
\begin{tabular}{ll}
\toprule
\textbf{Figure Component} & \textbf{Formal Notation} \\
\midrule

\textbf{Reasoning Agent (LLM)} & $M$ \\

\textbf{Obfuscated Context} & $\hat{s} \in \hat{\mathcal{S}}$ \\
\textbf{Obfuscation Membrane} & $\Pi : \mathcal{S} \rightarrow \hat{\mathcal{S}}$ \\

\textbf{Intent / Proposed Action} & $i \in \mathcal{I}$ \\
\textbf{Justification} & $J$ \\

\textbf{Sovereign Control Plane} & $(E, \mathcal{X})$ \\

\textbf{Policy Evaluation} & $E(i, s, \mathcal{P})$ \\
\textbf{System State (True Context)} & $s \in \mathcal{S}$ \\

\textbf{Scoped Execution} & $\mathcal{X}(i, s)$ \\
\textbf{Executed Action} & $a_{\mathrm{exec}}$ \\

\textbf{Context Snapshot} & $c_t$ \\
\textbf{Evidence Chain (Ledger)} & $\mathcal{C}_t$ \\

\textbf{Feedback Loop (State Update)} & $s_{t+1} = f(s_t, a_{\mathrm{exec}, t})$ \\

\bottomrule
\end{tabular}
\caption{Mapping between Figure~2 components and formal notation in the SAL model.}
\label{tab:figure_mapping}
\end{table}

\end{document}